\documentclass[floats,aps,article,onecolumn,showpacs]{revtex4}
\usepackage{graphicx}
\usepackage{amsmath}
\begin{document}
\newcommand {\nc} {\newcommand}
\nc {\beq} {\begin{eqnarray}} \nc {\eol} {\nonumber \\} \nc {\eeq}
{\end{eqnarray}} \nc {\eeqn} [1] {\label{#1} \end{eqnarray}} \nc
{\eoln} [1] {\label{#1} \\} \nc {\ve} [1] {\mbox{\boldmath $#1$}}
\nc {\rref} [1] {(\ref{#1})} \nc {\Eq} [1] {Eq.~(\ref{#1})} \nc
{\re} [1] {Ref.~\cite{#1}} \nc {\dem} {\mbox{$\frac{1}{2}$}} \nc
{\arrow} [2] {\mbox{$\mathop{\rightarrow}\limits_{#1 \rightarrow
#2}$}}

\author {E.M. Tursunov}
\affiliation {Institute of Nuclear Physics, Uzbekistan Academy of Sciences, \\
100214, Ulugbek, Tashkent, Uzbekistan}
\author {S. Krewald}
\affiliation{Institute fuer Kernhysik, Forschungszentrum Juelich, \\
52425, Juelich, Germany}

\title{Spectrum of the excited $N^*$ and $\Delta^*$ baryons in a relativistic
chiral quark model}

\date{\today}

\begin{abstract}
 \par The spectrum of the SU(2) flavor baryons
 is studied in the frame of a relativistic chiral quark potential model
 based on the one-pion and one-gluon exchange mechanisms.
 It is argued that the $N^*$ and $\Delta^*$ resonances appearing in the $\pi N$
scattering data and strongly coupled to the $\pi N$ channel are
identified with the orbital configurations $(1S_{1/2})^2(nlj)$ with
a single valence quark in the excited state $(nlj)$. With the
obtained selection rules based on the "chiral constraint",
 we show that it is possible to construct a
schematic periodic table of baryon resonances, consistent with the
experimental data and yielding no "missing resonances".
\par A new original method for the center of mass correction problem
of the zero-order three-quark core energy values of the excited
baryon resonances is suggested, which is based on the separation of
the three-quark Dirac Hamiltonian into the parts, corresponding to
the Jacobi coordinates.
\par The numerical estimations for the energy positions
 of the Nucleon and Delta baryons (up to and including F-wave
 $N^*$ and $\Delta^*$ resonances), obtained within the field-theoretical framework
by using time ordered perturbation theory, yield an overall good
description of the experimental data at the level of the relativized
Constituent quark model of S. Capstick and W. Roberts without any
fitting parameters. The only free parameter of the linear
confinement potential was fitted previously by Th. Gutsche to
reproduce the axial charge of the Nucleon. The ground state
$\Delta(1232)$ is well reproduced. However, Nucleon ground state and
most of the radially excited baryon resonances (including Roper) are
overestimated. Contrary, the first band of the orbitally excited
baryon resonances with a negative parity are underestimated. At the
same time, the second band of the orbitally excited  $\Delta*$
states with the negative parity are mostly overestimated, while the
$N^*$ states are close to the experimental boxes.
\par The theoretical estimations of the energy levels for the positive
parity baryon resonances with J=5/2, 7/2 are close to the
experimental data.
\par At higher energies, where the experimental data are poor, we can
extend our model schematically and  predict an existence of seven
$N^*$ and four $\Delta^*$ new states with larger spin values.
\end{abstract}
\pacs{11.10.Ef,12.39.Fe,12.39.Ki}

\maketitle

\section{Introduction}

\par
Presently,   several experimental collaborations study the production and decay of excited baryons
motivated by open questions concerning the origin of the hadronic masses\cite{rev}. The successes
of lattice Quantum Chromodynamics(QCD) in describing the ground state hadrons confirm the
QCD Lagrangian in the non-perturbative regime of the strong interaction\cite{fodorHoe}. First studies
of the masses of excited baryons are available.  While the lattice simulations  are numerically very
involved, there are simpler empirical rules which work amazingly well: the Forkel-Klempt mass
formula  reproduces the known N$^*$ and $\Delta^*$ masses with  three parameters\cite{rev}.
The question arises whether the lattice results can be interpreted by simpler models, such as  flux tubes or constituent quarks.
The Constitutent Quark Models(CQM) are the oldest approaches to baryon spectroscopy and have evolved
into three major subspecies, based on
 the Goldstone-boson exchange (GBE)
\cite{glo98}, the one-gluon exchange (OGE) \cite{cap86,cap00} or
(and) instanton induced exchange (IIE) \cite{met01} mechanisms
between ( non-) relativistic constituent quarks. In the present approach, we start from  relativistic chiral quark  models
\cite{thomas80,thomas08,saito84,saito08,waka10} which respect  the chiral symmetry.
 There are no studies of the excited baryon spectrum within these
 approaches in the literature.
\par In \cite{tur05,tur09,tur10} we have developed a relativistic
chiral quark model for the lower excitation spectrum of the nucleon
and delta. The splitting of the Roper resonance from the N(939) was
reproduced with a reasonable accuracy. The model was tested firstly
in Ref. \cite{oset84} for the study of the Nucleon charge form
factors, then in Ref. \cite{gutsche87,gutsche89} for the study of
the nucleon properties such as mass, charge radius, magnetic moment,
axial charge and reasonable agreement with the experimental values
was obtained.
\par The model is based on an effective chiral Lagrangian. Quark wave
 function is obtained from the solution
of the Dirac equation with a Cornell type potential containing a
linear confining term and a Coulomb part due-to short range gluon
field correlations. All the model parameters of the model, except
one are fixed from the Lattice study of previous authors
\cite{kawa11,green03}. The only free parameter of the model is the
so-called "mass term" in the confinement potential, which was fitted
in Ref.\cite{gutsche87} to reproduce the axial charge of the
Nucleon. The calculations are done at one loop or at order of
accuracy $o(1/f_{\pi}^2, \alpha_s)$.

\par The aim of present  paper is to extend the relativistic chiral quark
 model to the higher excitation spectrum of SU(2) flavor baryons.
Firstly we want to check, whether the relativistic chiral quark
model can help to understand the systematics of the excited Nucleon
and Delta states and an orbital structure of each baryon state.
Based on selection rules obtained from one-pion and one gluon
exchange mechanisms between valence quarks, below we will show that
it is possible to construct a periodic table, where each excited
Nucleon or Delta state can be identified with an orbital
configuration $(1S)^2(nlj)$ with a single radially or/and orbitally
excited valence quark.
\par Secondly we will estimate the excited Nucleon and Delta
spectrum in the present model with taking into account second-order
perturbative corrections due-to the pion and color-magnetic gluon
fields and compare with the experimental data.
  \par The relevant suggestion is that the results
 of our study can be reproduced in any chiral quark model describing the
baryons as bound states of three valence quarks with a Dirac
two-component structure and surrounded by the cloud of $\pi$-mesons,
as required by the chiral symmetry \cite{gell60}.
\par In Section 2 we give the main
 formalism of the model. The numerical results are presented in Section 3,
 and final conclusions are given in Section 4.
\section{Model}
\subsection{Basis formalism}
\par The effective Lagrangian of the model
${\cal L}(x)$ (see \cite{wei84,gutsche89}) contains the quark core
part ${\cal L}_Q(x)$, the quark-pion
 ${\cal L}_I^{(q\pi)}(x)$ and the quark-gluon ${\cal L}_I^{(qg)}(x) $
 interaction terms, and the kinetic parts for the pion ${\cal L}_{\pi}(x)$ and
gluon ${\cal L}_{g}(x)$ fields:
\begin{eqnarray}
\nonumber {\cal L}(x) = {\cal L}_Q(x) + {\cal L}_I^{(q\pi)}(x) +
{\cal
  L}_I^{(qg)}(x)+ {\cal L}_{\pi}(x) +  {\cal L}_{g}(x)  \\
 \nonumber
 = \bar\psi(x)[i\not\!\partial -S(r)-\gamma^0V(r)]\psi(x) - 1/f_{\pi}
 \bar\psi[S(r) i \gamma^5 \tau^i \phi_i]\psi- \\
 -g_s \bar\psi A_{\mu}^a\gamma^{\mu}\frac{\lambda^a}{2} \psi +
  \frac{1}{2}(\!\partial_{\mu}\phi_i)^2 - \frac{1}{2}m_{\pi}^2
  \phi_i^2-{1\over 4}G^a_{\mu\nu} G_a^{\mu\nu}.
\end{eqnarray}
Here, $\psi(x)$, $\phi_i, i=1,2,3$ and $A_{\mu}^a$ are the quark,
pion and gluon fields, respectively. The matrices $\tau^i (i=1,2,3)$
and $\lambda^a (a=1,...,8)$ are the isospin and color matrices,
correspondingly. The pion decay constant $f_\pi=$93 MeV. In the
model, the chiral symmetry violated through the quark confinement
 mechanism is restored with the help of the linearized $\sigma$-model.
  The mass term for the pion field is introduced in order to satisfy the PCAC
theorem \cite{cole68}, which is consistent with the
 Goldberger-Treiman relation.
\par  We use the Cornell type potential in the Dirac equation for the single quark
states in accordance with the Lattice QCD theory.
 The scalar part of the static confinement potential is given by
\begin{equation}\label{linpot}
S(r)=cr+m
\end{equation}
where $c$ and $m$ are constants. The strength parameter $c$ of the
confinement potential is defined from the Lattice study
\cite{kawa11}, while  $m$ is the only free parameter of the model
which can be fitted to reproduce the axial charge $g_A$ of the
proton (and the $\pi NN$ coupling constant via the
Goldberger-Treiman relation).

\par At short distances, transverse fluctuations of the string are dominating
\cite{lus81}, with an indication that they transform like the time
component of the Lorentz vector. They are given by a Coulomb type
vector potential (the so called Luscher term) as
\begin{equation}
\label{Coulomb}
 V(r)=-\alpha/r
\end{equation}
where $\alpha=\pi/12$ is defined from the QCD Lattice study
\cite{green03}.

\par The quark fields are obtained from solving the
Dirac equation with the corresponding scalar plus vector potentials
\begin{equation} \label{Dirac}
[i\gamma^{\mu}\partial_{\mu} -S(r)-\gamma^0V(r)]\psi(x)=0
\end{equation}
The respective positive and negative energy eigenstates as solutions
to the Dirac equation with a spherically symmetric mean field, are
given in a general form as
\begin{eqnarray} \label{Gaussian_Ansatz}
 u_{\alpha}(x) \, = \,
\left(
\begin{array}{c}
g^+_{N\kappa }(r) \\
-i f^+_{N\kappa }(r) \,\vec{\sigma}\hat{\vec x} \\
\end{array}
\right) \, {\cal Y}_{\kappa}^{m_j}(\hat{\vec x}) \,\chi_{m_t} \,
\chi_{m_c} \, exp(-iE_{\alpha}t)
\end{eqnarray}

\begin{eqnarray}
 v_{\beta}(x) \, = \,
\left(
\begin{array}{c}
g^-_{N\kappa}(r) \\
-i f^-_{N\kappa}(r) \,\vec{\sigma}\hat{\vec x} \\
\end{array}
\right) \, {\cal Y}_{\kappa}^{m_j}(\hat{\vec x}) \,\chi_{m_t} \,
\chi_{m_c} \, exp(+iE_{\beta}t)
\end{eqnarray}
The quark and anti-quark eigenstates $u$ and $v$ are labeled by the
radial, angular, azimuthal, isospin and color quantum numbers $N,\,
\kappa,\, m_j,\, m_t$ and $m_c$, which are collectively denoted by
$\alpha$ and $\beta$, respectively. The spin-angular part of the
quark field operators
\begin{equation}
{\cal Y}_{\kappa}^{m_j}(\hat{\vec x})\,=\,[Y_l(\hat{\vec x})\otimes
\chi_{1/2}]_{jm_j} \, \, j=|\kappa|-1/2.
\end{equation}
The quark fields $\psi$ are expanded over the basis of positive and
negative energy eigenstates as
\begin{equation}
\psi(x)=\sum \limits_{\alpha} u_{\alpha}(x)b_{\alpha} +\sum
\limits_{\beta} v_{\beta}(x)d^{\dag}_{\beta} .
\end{equation}
The expansion coefficients $b_{\alpha}$ and $d^{\dag}_{\beta}$ are
operators, which annihilate a quark and create an anti-quark in the
orbits $\alpha$ and $\beta$, respectively.
\par The free pion field operator is expanded over plane wave solutions as
\begin{equation}
\phi_j(x)=(2\pi)^{-3/2}\,
\int\frac{d^3k}{(2\omega_k)^{1/2}}[a_{j{\bf
k}}exp(-ikx)+a^{\dag}_{j{\bf k}}exp(ikx)]
\end{equation}
with the usual destruction and creation operators $a_{j{\bf k}}$ and
$a^{\dag}_{j{\bf k}}$ respectively. The pion energy is defined as \\
$\omega_k \,=\, \sqrt{k^2+m_{\pi}^2}. $ The free gluon field
operators is expanded in the same way.

\par In denoting the three-quark vacuum state by $ |0> $, the corresponding
noninteracting many-body quark Green's function (propagator) of the
quark field is given by the customary vacuum Feynman propagator for
a binding potential \cite{fet71}:
\begin{equation}
iG(x,x')\,=\, iG^F(x,x')\,=\,<0|T\{\psi(x) \bar\psi(x')\}|0>\,=\,
\sum \limits_{\alpha} u_{\alpha}(x)\bar u_{\alpha}(x')\theta(t-t') +
\sum \limits_{\beta} v_{\beta}(x)\bar v_{\beta}(x')\theta(t'-t)
\end{equation}
Since the three-quark vacuum state $|0>$ does not contain any pion
or gluon, the pion and gluon Green's functions are given by the
usual free Feynman propagator for a boson field:
\begin{equation}
i\Delta_{ij}(x-x')\,=\, <0|T\{\phi_i(x) \bar\phi_j(x')\}|0>\,=\,
i\delta_{ij}\int\frac{d^4k}{(2\pi)^4}\frac{1}{k^2-m_{\pi}^2+i\epsilon}
\,exp[-ik(x-x')] \, ,
\end{equation}

\begin{equation}
i\Delta^{(\mu\nu)}_{ab}(x-x')\,=\, <0|T\{ A^a_\mu (x)
A^b_{\nu}(x')\}|0>\,=\,
 i\delta_{ab}g^{\mu \nu}\int\frac{d^4k}{(2\pi)^4}\frac{1}{k^2+i\epsilon}
\,exp[-ik(x-x')] \,,
\end{equation}
(in the Coulomb gauge), where $g^{\mu\nu}=\delta_{\mu\nu}
g^{\mu\mu}$, $g^{00}=-g^{11}=-g^{22}=-g^{33}=1$ .
\par On the basis of the effective Lagrangian and using the time-ordered perturbation theory
within the frame of many-body quantum field theory \cite{fet71} we
can develop the calculation scheme for the excitation spectrum of
the Nucleon and Delta.
 At zero-th order the quark
core result ($E_Q$) is obtained by solving Eq.(\ref{Dirac}) for the
single quark system numerically by using the harmonic oscillator
basis. Since we work in the independent particle model, we assume
that the bare three-quark state of the $SU(2)$-flavor baryons
corresponds to the structure $(1S_{1/2})^2(nlj)$ with a single
excited valence quark in the non-relativistic spectroscopic
notation. Below, on the basis of the one-pion and one-gluon exchange
mechanisms we will argue that such a configuration of the three
valence quarks is identified with the baryon resonances decaying
strongly  into the $\pi + N$ channel. And contrary, the baryon
states with more than one valence quarks in excited orbits do not
have a strong coupling into this channel. In other words, all baryon
resonances appearing in the $\pi N$ scattering data can be
identified with the above orbital configuration containing a single
excited valence quark. This is why we fix the excited baryon
configuration as $(1S_{1/2})^2(nlj)$. The corresponding quark core
energy is evaluated as the sum of single quark energies with:
\begin{equation}
E_Q=2E(1S_{1/2}) + E(nlj)
\label{Core}
\end{equation}

 \par The second order perturbative corrections to the energy spectrum of the
 SU(2) baryons due to the pion ($\Delta E^{(\pi)}$) and  gluon ( $\Delta
 E^{(g)}$) fields are calculated on the basis of the Gell-Mann and Low theorem :
 \begin{eqnarray}\label{Energy_shift}
\hspace*{-.8cm} \Delta E=<\Phi_0| \, \sum\limits_{n=1}^{\infty}
\frac{(-i)^n}{n!} \, \int \, i\delta(t_1) \, d^4x_1 \ldots d^4x_n \,
T[{\cal H}_I(x_1) \ldots {\cal H}_I(x_n)] \, |\Phi_0>_{c}
\end{eqnarray}
with $n=2$, where the relevant quark-pion and quark-gluon
interaction Hamiltonian densities are
\begin{eqnarray}
{\cal H}_I^{(q\pi)}(x)= \frac{i}{f_{\pi}}\bar\psi(x)\gamma^5
\vec\tau\vec\phi(x)S(r)\psi(x),
\end{eqnarray}
\begin{eqnarray}
{\cal H}_I^{(qg)}(x)= g_s\bar\psi(x)A_{\mu}^a(x)\gamma^{\mu}
\frac{\lambda^a}{2}\psi(x)
\end{eqnarray}
The stationary bare three-quark state $|\Phi_0>$ is constructed from
the vacuum state using the usual creation operators:
\begin{equation}
|\Phi_0>_{\alpha\beta\gamma}=b_{\alpha}^+b_{\beta}^+b_{\gamma}^+|0>,
\end{equation}
where $\alpha, \beta$ and $ \gamma$ represent the quantum numbers of
the single quark states, which are coupled to the respective baryon
configuration. The energy shift of Eq.(\ref{Energy_shift}) is
evaluated up to second order in the quark-pion and quark-gluon
interaction, and generates self-energy and exchange diagrams
contributions. In the self-energy diagrams a single pion or gluon is
emitted and absorbed by the same valence quark, which however can be
excited to an intermediate quark or anti-quark state. In the second
order exchange diagrams a single pion or gluon, emitted by a valence
quark is absorbed by another valence quark of the SU(2) baryon.

\subsection{Center of mass corrections for the ground state N and $\Delta$}

The result for $E_Q$ in Eq.(\ref{Core}) contains an essential
spurious  contribution of the center of mass motion to the energy of
the baryons. A covariant way of the separation of the CM motion is
possible in non-relativistic models. In the nonrelativistic three
nucleon system the energy is reduced by factor about 1/3 after the
separation of the CM. At the same time different approaches are
being used in relativistic mean field models.

For the ground state nucleon and delta baryons we use the
development of the Ref. \cite{shimizu}, where three different
approximations have been used, which estimate corrections for the
center of mass motion: the $R=0$ \cite{lu98}, $P=0$ \cite{teg82} and
LHO \cite{wil89} methods. In all three methods the baryon wave
function is rewritten in the Jacobi coordinates in the center of
mass system as $\Phi_B(\vec{r}, \vec{\rho},\vec{R})$, where
$\vec{r}$, $\vec{\rho}$ and $\vec{R}$ are relative coordinates
between the two valence quarks, between 3-valence quark and the
center of mass of the 1+2 quarks, and the center of mass of the all
three quarks, respectively:

\begin{eqnarray}
\nonumber
\vec{r}=\vec{r_1}-\vec{r_2} \\
\nonumber
\vec{\rho}=(\vec{r_1}+\vec{r_2})/2- \vec{r_3} \\
\vec{R}= (\vec{r_1}+\vec{r_2}+ \vec{r_3})/3
\end{eqnarray}

The initial baryon wave function $\Phi(\vec{r_1},
\vec{r_2},\vec{r_3})$ expanded in the oscillator basis states are
transformed to the Jacobi coordinates by using the Moshinsky
transformation (see Ref \cite{shimizu} for details).

In the R=0 method the baryon wave function in the CM system is
multiplied by the plane wave of the CM motion:
 \beq
 \Phi_R(\vec{r_1},
\vec{r_2},\vec{r_3},\vec{P})=N_R \,\, exp(i\vec{P} \cdot
\vec{R})\Phi_B(\vec{r}, \vec{\rho},\vec{R}=0).
 \eeq
The second P=0 method is based on the Fourier transformation of the
baryon wave function:
\beq
 \Phi_P(\vec{r_1},\vec{r_2},\vec{r_3},\vec{P})=N_P \,\, exp(i\vec{P} \cdot \vec{R})
 \int exp(-i\vec{P} \cdot \vec{R'})\Phi_B(\vec{r},\vec{\rho},\vec{R'}) d \vec{R'}.
 \eeq
The lowest harmonic oscillator (LHO) method is based on the
projection of the baryon wave function on the lowest harmonic
oscillator state: \beq
 \Phi_{LHO}(\vec{r_1},\vec{r_2},\vec{r_3},\vec{P})=N_{LHO} \, \, exp(i\vec{P} \cdot \vec{R})
 \int R_{0s}(\vec{R'})\Phi_B(\vec{r}, \vec{\rho},\vec{R'}) d \vec{R'}.
 \eeq
The factors $N_R$, $N_P$ and $N_{LHO}$ differ each from other and
are found from the normalization conditions: \beq
 <\Phi(\vec{r_1},\vec{r_2},\vec{r_3},\vec{P})|
 \Phi(\vec{r_1},\vec{r_2},\vec{r_3},\vec{P'})> = (2\pi)^3\delta(\vec{P}-\vec{P'}).
 \eeq
In all of the three methods, the average kinetic energy and mass
terms of the three-body system are estimated by using angular
momentum algebra and numerical methods (see Ref.\cite{shimizu} for
details).

\subsection{Center of mass corrections for the excited states N* and
$\Delta$*}\label{CMmodel}

\par For the excited nucleon and delta states with fixed orbital
configuration $(1S)^2(nlj)$, the Moshinsky transformation is not
applicable due-to two-component structure of the valence quark wave
functions. An original  new approach to the center of mass
correction problem is based on the separation of the total
three-quark core Dirac Hamiltonian with the scalar and vector mean
field potentials
 \beq \label{hamilcm}
\hat{H}= \sum_{i=1}^3 [{\vec{\alpha_i} \vec{p_i}
+S(\vec{r_i}-\vec{R})\beta_i + V(\vec{r_i}-\vec{R}) }]
 \eeq
into two parts corresponding to the relative motion and center of
mass motion, respectively. In this way one can estimate the
zero-order quark-core energy for the baryon states with fixed
orbital configurations $(1S)^2(nlj)$ free off the center of mass
motion by solving the corresponding equation. At the zero order the
energy values of all baryon states with fixed orbital configuration
degenerate. This means that one can estimate the zero-order energy
values of baryon states with the fixed orbital configuration
$(1S)^2(nlj)$, assuming that the two S-quarks are in the $^1S_0$
singlet scalar diquark state.
\par The kinetic energy term can be
rewritten easily as:
\begin{eqnarray}
\nonumber
 \hat{H_0}=   \hat H_{R,0} + \hat H_{rel,0},  \\
 \nonumber
\hat H_{R,0}= \frac{\vec{\alpha_1}+\vec{\alpha_2}+\vec{\alpha_3}}{3} \vec{P_R}, \\
\hat H_{rel,0}= (\vec{\alpha_1}-\vec{\alpha_2})  \vec{P_r} +
(\frac{\vec{\alpha_1}+\vec{\alpha_2}}{2}-\vec{\alpha_3})\vec{P_{\rho}}.
\end{eqnarray}
\par First we study the center of mass motion problem for the scalar-vector mean-field
potential of the oscillator form
 \begin{eqnarray} \label{inter}
 \nonumber
 \hat H_{int}=\sum_{i=1}^{3}[V_1(\vec{r_i}-\vec{R})\beta_i + V_2(\vec{r}_i-\vec{R})], \\
  V_k(\vec{r_i})=c_k r_i^2+\mu_k,  k=1,2.
 \end{eqnarray}
 In this case the interaction part of the three-quark core Hamiltonian
 can be exactly separated in  Jacobi coordinates as
 \begin{eqnarray} \label{modpot}
 \nonumber
 \hat H_{int}=V_r +V_{\rho} \\
 \nonumber
 V_r= 1/2(c_1 \beta_r + c_2) r^2+ 2(\mu_1\beta_r +\mu_2) \\
V_{\rho}= 2/3 (c_1 \beta_{\rho}+ c_2) \rho^2
+\mu_1\beta_{\rho}+\mu_2 ,
 \end{eqnarray}
where we introduced the Dirac matrices $\beta_r$ and $\beta_{\rho}$
corresponding to the Jacobi coordinates $r$ and $\rho$,
respectively. In consistence with the above assumption, that the two
S-quarks are in the singlet $^1S_0$ state and combining the kinetic
and interaction parts of the relative motion Hamiltonian for the
case of the oscillator scalar-vector mean field potentials we can
write down:
 \beq
 \hat H_{rel}=\hat H_r+ \hat H_{\rho},
\eeq where the Hamiltonian $\hat H_r$ corresponds to the singlet
diquark relative motion, and the Hamiltonian $\hat H_{\rho}$ is
related to the single excited valence quark motion with the modified
potentials:

\begin{eqnarray} \label{hamiljacobi}
  \nonumber
 \hat H_r= (\vec{\alpha}_1-\vec{\alpha}_2) \vec{P}_r +V_r  \\
\hat H_{\rho} = -\vec{\alpha}_3\vec{P}_{\rho} + V_{\rho}.
\end{eqnarray}
The two-body Dirac equation
 \beq \label{twobody}
 \hat H_r \Psi (\vec r)= E_r \Psi(\vec r)
\eeq can be solved in the same way as the single particle Dirac
equation with the only difference that the lower component of the
two-body Dirac wave function differs from the upper component by the
both spin and orbital momentum. This result is a consequence of the
relation:
 \begin{eqnarray}
 (\vec \sigma_1-\vec \sigma_2)\hat{\vec{r}} \,\,
{\cal Y}_{l,S}^{jm_j}(\hat{\vec r})  = -2 \sqrt{3} \sum_h (2h+1)
 \left\{
\begin{array}{ccc}
 \frac{1}{2} &  \frac{1}{2} & S \\
 l   &  j &  h \\
\end{array}
\right \}
\left\{
\begin{array}{ccc}
 \frac{1}{2} &  \frac{1}{2} & (S\pm 1)\\
 (l\pm 1)  &  j &  h \\
\end{array}
\right \} {\cal Y}_{l\pm 1,S\pm 1}^{jm_j}(\hat{\vec r}) ,
\end{eqnarray}
where
\beq
    {\cal Y}_{l S}^{j m_j}(\hat{\vec r})= [Y_{l}(\hat{\vec r})\otimes
    \chi_{S}(1,2)]_{jm_j},
\eeq

which shows us the correct form for the two-body Dirac bound state
wave function to be as
 \beq
 \Psi (\vec r)=
\left(
\begin{array}{c}
g_{N,l }(r) {\cal Y}_{l,S}^{jm_j}(\hat{\vec r})\\
i f_{N,l\pm 1}(r) {\cal Y}_{l\pm 1,S\pm 1}^{m_j}(\hat{\vec r})  \\
\end{array}
\right).
 \eeq
The radial wave functions $g_{N,l}(r)$ and $f_{N,l}(r)$ are expanded
over the oscillator basis states as was done for the single quark
wave function. For the scalar diquark in the ground state the upper
and lower components of the two-body Dirac wave function present the
$^1S_0$ and $^3P_0$ waves, respectively. The estimated energy value
of the Eq.(\ref{twobody}) together with the solution of the
single-quark Dirac equation
\begin{eqnarray}\label{single}
  \hat H_{\rho} \Psi(\vec \rho) = E_{\rho} \Psi(\vec \rho)
\end{eqnarray}
with the modified potential Eq. (\ref{modpot}) yield us the quark
core results of the energy value for the excited baryon states with
the fixed orbital configuration $(1S)^2(nlj)$
 \beq E_0=E_r+E_{\rho}
\eeq free off the center of mass contribution. Thus, we found a way
to separate the center of mass motion of the three-quark system
bound by the scalar-vector mean-field oscillator potentials.
    \par Now we return to the Eq. (\ref{hamilcm}) with the interaction Hamiltonian
\begin{eqnarray} \label{realpot}
 \hat H_{int}=\sum_{i=1}^{3}[S(\vec{r_i}-\vec{R})\beta_i + V(\vec{r}_i-\vec{R})], \\
\end{eqnarray}
with the linear scalar $S(r)=cr+m$ (see  Eq.(\ref{linpot})) and
Coulomb-like vector $V(r)=\alpha/r$ (see Eq. (\ref{Coulomb}))
mean-field potentials. For these potentials, unlike scalar-vector
mean-field potentials, the separation of the interaction Hamiltonian
on the potentials $V_r$ and $V_{\rho}$, dependent on the Jacobi
coordinates $r$ and $\rho$, respectively, is a strong task. For the
confinement potential in the Jacobi coordinates we have an expansion
over multipols
\begin{eqnarray}
 \hat S(\vec r,\vec \rho)=  \sum_{i=1}^{3} S(\vec{r_i}-\vec{R})
 =2 c \sum_{l=0,2,...} \frac{(\rho/3)^l}{(r/2)^{l+1}} \, \,
 \left[ \frac{\rho^2/9}{2l+3} -\frac{r^2/4}{2l-1} \right] P_l(cos(\vec r \,\hat ,\vec
 \rho)) +\frac{2}{3}c\rho + 3m,
\end{eqnarray}
where $P_l(cos(\vec r \,\hat ,\vec \rho))$ are the Legandre
polynomials. The Coulomb-like potential is transformed in the same
way into the Jacobi coordinates as
\begin{eqnarray}
 \hat V (\vec r,\vec \rho)=\sum_{i=1}^{3} V(\vec{r}_i-\vec{R})=
 \frac{4 \alpha}{r} \sum_{l=0,2,...} \left (\frac{\rho/3}{r/2} \right )^l \, \,
  P_l(cos(\vec r \,\hat ,\vec \rho))  + \frac{3 \alpha}{2\rho}.
\end{eqnarray}
The above equations are valid for  $\rho/3 < r/2$. In the rest area
 these variables must be interchanged. Thus, in the Jacobi
coordinates we come to the situation, when the original linear
confinement and Coulomb-like potentials depend on the angle between
the Jacobi coordinate-vectors $\vec r$ and $\vec \rho$. However, at
first approximation when keeping the main multipols, these
potentials can be written as
\begin{eqnarray}
\nonumber
 \hat S(\vec r,\vec \rho) \approx c r + \frac{2}{3}c\rho + 3m, \\
 \hat V (\vec r,\vec \rho) \approx \frac{4 \alpha} {r} + \frac{3 \alpha}{2\rho}.
\end{eqnarray}
\par On the basis of the last approximation we divide the
confinement and Coulomb potential terms in the Jacobi coordinates
into two parts according to the Eq. (\ref{hamiljacobi}) and
corresponding to the scalar diquark plus the modified  single quark
Hamiltonians. The first test calculations can be done for the
separation
\begin{eqnarray}\label{effecpot}
\nonumber
 \hat S(\vec r) =  c r +  2m, \, \, \, \, \,  \hat S(\vec \rho) =\frac{2}{3}c\rho + m, \\
 \hat V (\vec r) =\frac{4 \alpha} {r}, \, \, \, \, \, \,\,\,\,\, \hat V (\vec \rho) =  \frac{3 \alpha}{2\rho}.
\end{eqnarray}
With the help of these separated effective potentials, we can
estimate the energy values of the scalar diquark and the single
valence quark, which give us the three-quark core energy value at
the zero order, free off the center of mass motion contribution.
However, it is important to note that the effective potentials for
the diquark in the Eq. (\ref{effecpot}) in fact coincide completely
with the two-body potentials, derived from the original single-quark
confinement scalar  $S(r)=cr+m$ (see  Eq.(\ref{linpot})) and
Coulomb-like vector $V(r)=\alpha/r$ (see Eq. (\ref{Coulomb}))
mean-field potentials. This means that these effective potentials
are exact for the free diquark system, but not for the bound diquark
inside the baryon. In reality, the diquarks are bound with an
additional valence quark and have lighter mass than free diquarks.
This is why in further we slightly increase the attraction in the
diquark effective potentials by fitting them to reproduce the
quark-core energy value of the ground state Nucleon, estimated by
one of the methods, described in the previous section. Then we can
employ the effective potentials for the solution of the Eq.
(\ref{twobody}) and Eq. (\ref{single}) to estimate the quark core
energy values of the excited $N^*$ and $\Delta^*$ resonances by
using the developed in the present Section method.

\subsection{ Self-energy diagrams contribution}
\par The self-energy terms contain contribution both from
intermediate quark $(E>0)$ and anti-quark $(E<0)$ states. These
diagrams describe the processe when a pion or gluon is emitted and
absorbed by the same valence quark which can be excited to the
intermediate quark or anti-quark states.
\par
\begin{figure}[tbh]
\begin{center}
\includegraphics[width=15cm]{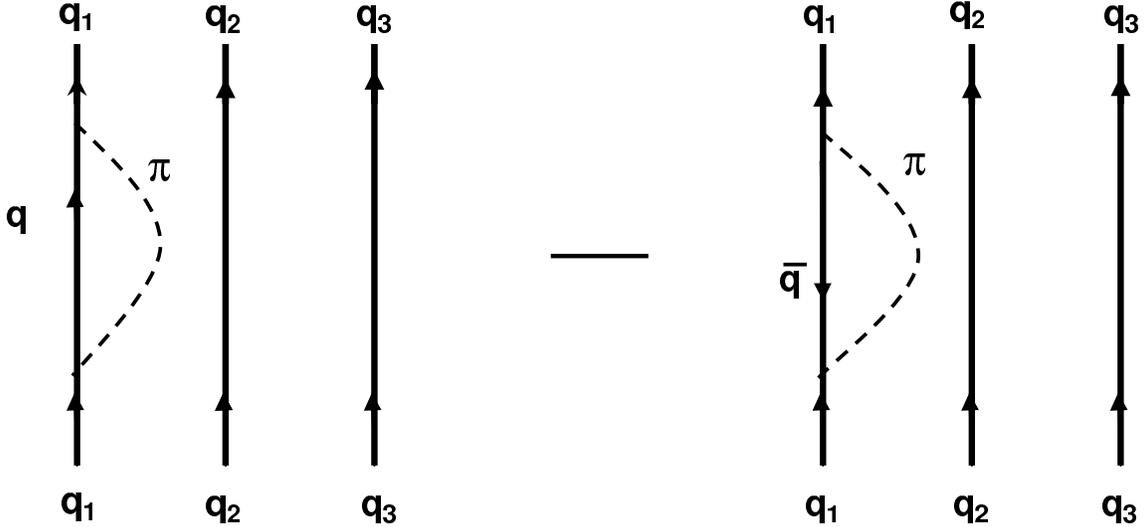}
\end{center}
\caption{Second order self energy diagrams induced by $\pi-$meson
fields \label{Fig1a}}
\end{figure}

 The pion part of the self energy term (pion cloud contribution)
 (see Fig.\ref{Fig1a} ) is evaluated as
\begin{eqnarray}
\Delta
E_{s.e.}^{(\pi)}=-\frac{1}{2f_{\pi}^2}\sum\limits_{a=1}^{3}\sum\limits_{\alpha
  ' \leq \alpha_F} \int
\frac{d^3\vec p}{(2\pi)^3p_0} \biggl\{
\sum\limits_{\alpha}\frac{V_{\alpha
      \alpha ' }^{a+}(\vec p)V_{\alpha \alpha ' }^{a}(\vec
        p)}{E_{\alpha}-E_{\alpha '}+p_0}-
 \sum\limits_{\beta}\frac{V_{\beta \alpha '}^{a+}(\vec p) V_{\beta
 \alpha '}^{a}(\vec p)}{E_{\beta}+E_{\alpha '}+p_0}\biggr\},
\end{eqnarray}
with $ p_0^2=\vec p^2 + m_{\pi}^2$. The $q-q-\pi$ transition form
factors are defined as:
\begin{eqnarray}
V_{\alpha\alpha'}^a(\vec p)=\int d^3 x \bar u_{\alpha}(\vec
x)\Gamma^a(\vec x)
u_{\alpha '}(\vec x) e^{-i\vec p \vec x} \\
V_{\beta \alpha'}^a(\vec p)=\int d^3 x \bar v_{\beta}(\vec
x)\Gamma^a(\vec x) u_{\alpha '}(\vec x) e^{-i\vec p \vec x}
\end{eqnarray}
\par The vertex function of the $\pi -q -q $ and $\pi -q -\bar q $
transition is
\begin{eqnarray}
\Gamma^a= S(r) \gamma^5 \tau^a I_c \, ,
\end{eqnarray}
where $I_c$ is the color unity matrix. The expression of the
$\pi-q-q$ transition form factor has been derived in
Ref.\cite{tur09}:
\begin{eqnarray}
\nonumber V_{\alpha\alpha'}^a(\vec p) & = & \sum_{l_n}
(-i)^{l_n+1}\int dr r^2\Big[g_{\alpha}(r)f_{\alpha'}(r)+
g_{\alpha'}(r)f_{\alpha}(r)\Big]S(r) j_{l_n}(pr) \\
 &  &  Y_{l_n}^{m_j'-m_j}(\hat{p}) {\cal F}(l^{\pm},l',l_n,j,j',m_j,m_j')
<m_t|\tau^a|m_t'><m_c|I_c|m_c'>.
\end{eqnarray}
The Hermitian conjunction of the transition form factor
\begin{eqnarray}
\nonumber V_{\alpha\alpha'}^{a+}(\vec p)& = & \sum_{l_n}
(i)^{l_n+1}\int dr r^2\Big[g_{\alpha}(r)f_{\alpha'}(r)+
g_{\alpha'}(r)f_{\alpha}(r)\Big]S(r) j_{l_n}(pr) \\
 &  &  Y_{l_n}^{(m_j'-m_j)*}(\hat{p}) {\cal
F}(l^{\pm},l',l_n,j,j',m_j,m_j') <m_t'|\tau^a|m_t><m_c'|I_c|m_c>.
\end{eqnarray}

After integration over the angular part in Eq. (17), the self-energy
diagrams contribution to the baryon spectrum induced by pion fields
 is evaluated as:
\begin{eqnarray}
\nonumber \Delta E_{s.e.}^{(\pi)}=-\frac{1}{16\pi^3f_{\pi}^2}\int
\frac{d p \, p^2}{p_0} \sum\limits_{\alpha ' \leq
\alpha_F}\sum\limits_{l_n} \biggl\{ \sum\limits_{\alpha}\frac{[\int
dr r^2 G_{\alpha \alpha
    '}(r)S(r)j_{l_n}(pr)]^2}{E_{\alpha}-E_{\alpha '}+p_0}
Q_{s.e.}(l,l',l_n,j,j') - \\
\sum\limits_{\beta}\frac{[\int dr r^2 G_{\beta \alpha
'}(r)S(r)j_{l_n}(pr)]^2}{E_{\beta}+E_{\alpha '}+p_0}
Q_{s.e.}(l,l',l_n,j,j') \biggr \} ,
\end{eqnarray}
where $j_{l_n}$ is the Bessel function. The radial overlap of the
single quark states with quantum numbers
$\alpha=(N,l,j,m_j,m_t,m_c)$ and $\alpha^{\prime}$ is defined as
\begin{eqnarray}
  G_{\alpha \alpha '}(r)=f_{\alpha}(r)g_{\alpha '}(r) + f_{\alpha '}(r)
g_{\alpha }(r)  .
\end{eqnarray}
The angular momentum coefficients $Q$ are evaluated for all SU(2)
baryons as
\begin{eqnarray}
Q_{s.e.}(l,l',l_n,j,j')= 12\pi [l^{\pm}][l_n][j] \biggl [
C^{l'0}_{l^{\pm}0l_n 0} W (j \frac{1}{2}l_nl'; l^{\pm}j')\biggr ]^2
\sum\limits_{m_j}\sum \limits_{m_j ' \leq \alpha_f} \biggl [
C^{j'm_j'}_{jm_jl_n(m_j'-m_j)} \biggr ]^2 ,
\end{eqnarray}
where $C$ and $W$ are the Clebsch-Gordan and Wigner coefficients,
respectively.
\par The gluon part of the second order self-energy diagrams (gluon cloud) contribution
is  estimated in a similar way as
\begin{eqnarray}
\Delta E_{s.e.}^{(g)}=\frac{g_s^2}{2}\sum\limits_{a}g_{\mu\nu}
\sum\limits_{\alpha ' \leq \alpha_F} \int\frac{d^3\vec p}{(2\pi)^3p}
\biggl\{ \sum\limits_{\alpha}\frac{V_{\alpha
      \alpha ' }^{a\mu+}(\vec p)V_{\alpha \alpha ' }^{a\nu}(\vec p)}
{E_{\alpha}-E_{\alpha '}+p}-
 \sum\limits_{\beta}\frac{V_{\beta \alpha '}^{a\mu+}(\vec p) V_{\beta
 \alpha '}^{a\nu}(\vec p)}{E_{\beta}+E_{\alpha '}+p}\biggr\},
\end{eqnarray}
where the transition form factor is evaluated with the corresponding
vertex matrix
\begin{eqnarray}
\Gamma_{\mu}^a=\gamma^{\mu}\frac{\lambda^a}{2}I_t
\end{eqnarray}
with the isospin unity matrix $I_t$.
\begin{equation}
V_{\alpha \alpha'}^{a\mu}(\vec{p})=\delta_{\mu 0}
\int{d^3x\bar{u}_{\alpha}(\vec{x})\frac{\lambda_a}{2}I_t
u_{\alpha'}(\vec{x})exp({-i\vec{p}\vec{x}})} + \delta_{\mu k}
\int{d^3x\bar{u}_{\alpha}(\vec{x})\frac{\lambda_a}{2}I_t
\hat{\alpha}_k u_{\alpha'}(\vec{x})exp({-i\vec{p}\vec{x}})}
\end{equation}
The last expression is convenient for the estimation of the exchange
diagrams.
\par For the self-energy diagrams we use an alternative
expression of the transition form-factors. Putting the quark wave
functions with further integration over the radial part of the
 spatial coordinate one can write for the transition form-factor
 next equation:
\begin{eqnarray}
\nonumber\
V_{\alpha\alpha'}^{a\mu}(\vec{p})=\sum_{l_{n}m_{n}}\sum_{LL'}
\sum_{m_{L}m_{L}'m_{s}m_{s}'}\Big(\frac{[L][l_n](4\pi)}
{[L']}\Big)^{\frac{1}{2}}(-i)^{l_{n}}
Y_{l_{n}m_{n}}(\hat{p})M_{m_{s}m_{s}'}^{\mu}C_{\mathrm{L0l_{n}0}}^{\mathrm{L'0}}
C_{\mathrm{Lm_{L}\frac{1}{2}m_{s}}}^{\mathrm{jm_{j}}}
C_{\mathrm{L'm_{L}'\frac{1}{2}m_{s}'}}^{\mathrm{j'm_{j}'}}
C_{\mathrm{Lm_{L}l_{n}m_{n}}}^{\mathrm{L'm_{L}'}}\cdot \\
\cdot\int{r^{2}R_{\mathrm{\mu_{LL'}}}^{\mathrm{\alpha\alpha'}}(r)j_{\mathrm{l_{n}}}(pr)dr}
<m_{t}|I_{t}|m_{t}'><m_{c}|\frac{\lambda_{a}}{2}|m_c'> ,
\end{eqnarray}
 where the spin transition matrices
\begin{displaymath}
 M_{m_s m_s'}^0 = \delta _{m_s m_s'},
\end{displaymath}
and
\begin{displaymath}
 M_{m_s m_s'}^k = \sum _{k' =\pm 1,0} h_{kk'} \Big[
 \delta_{k'1}\delta_{m_s 1/2}\delta_{m_s'(-1/2)} +
\delta_{k'(-1)}\delta_{m_s (-1/2)}\delta_{m_s'1/2} +
 2 m_s \delta_{k'0}\delta_{m_s m_s'} \Big]
\end{displaymath}
with the only nonzero expansion coefficients $
h_{1,+1}=h_{1,-1}=h_{3,0}=1$, and  $ h_{2,+1}= -h_{2,-1}= -i $ .
\par The radial functions are defined as
\begin{displaymath}
 R_{\mu_{LL'}}^{\alpha\alpha'}(r)=
\delta_{\mu,0}\delta_{Ll}\delta_{L'l'}
(g_{\alpha}g_{\alpha'}+f_{\alpha}f_{\alpha'} ) +i \delta_{\mu,k}
(\delta_{Ll}\delta_{L'l'^{\pm}} g_{\alpha} f_{\alpha'} -
\delta_{L'l'}\delta_{Ll^{\pm}} g_{\alpha'} f_{\alpha})
\end{displaymath}
\par
\begin{figure}[tbh]
\begin{center}
\includegraphics[width=15cm]{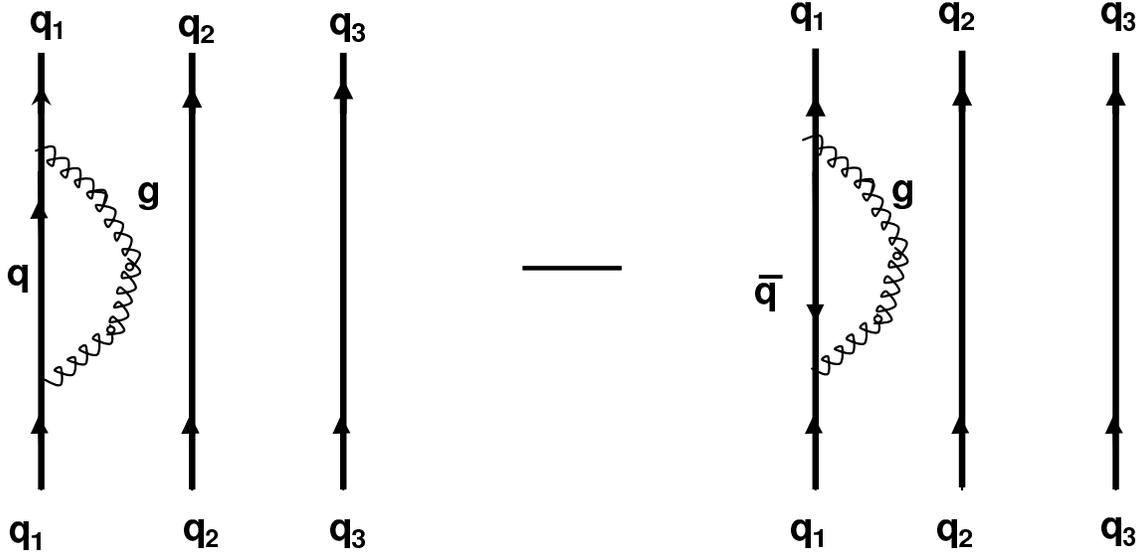}
\end{center}
\caption{Second order self energy diagrams induced by gluon fields
\label{Fig1b}}
\end{figure}

 The corresponding Feynman diagrams are given in Fig.\ref{Fig1b},
where the contribution from intermediate quark and anti-quark levels
have opposite signs.
\par  After evaluation of the transition form-factors and integration over angular variables,
the self-energy term induced by gluon fields can be written as a sum
of color-electric (Coulomb) and color-magnetic parts (see
Ref.\cite{tur09}):
\[
 \Delta E_{s.e.}^{(g)}=\frac{g_s^2}{3\pi^2} \sum\limits_{N'l'j'}\sum\limits_{(\alpha,\beta)}
\sum\limits_{LL'L^*L'^*l_n}[l_n]\left ( \frac{[L][L^*]}{[L'][L'^*]}
\right )^{1/2}C^{L'0}_{L0l_n0} C^{L'^*0}_{L^*0l_n0}
\]
\[
\biggl \{ \delta_{lLL^*}\delta_{l'L'L'^*}\delta_{l_nl}
{\cal A} ^{jj'm_jm_j'}_{LL'L^*L'^*l_n}  \\
\left [ \int \frac {[R_{\alpha \alpha'l_n}(p)+F_{\alpha
\alpha'l_n}(p)]^2} {E_{\alpha}-E_{\alpha '}+p}pdp - \int \frac
{[R_{\beta \alpha'l_n}(p)+F_{\beta \alpha'l_n}(p)]^2}
{E_{\beta}+E_{\alpha '}+p}pdp \right ]
\]
\[
 -\left [{\cal
B}^{jj'm_jm_j'}_{LL'L^*L'^*l_n} -{\cal
D}^{jj'm_jm_j'}_{LL'L^*L'^*l_n} + 2{\cal
E}^{jj'm_jm_j'}_{LL'L^*L'^*l_n} \right ]
\]
\begin{equation}
 \left[ \int\frac{dp \,p}{E_{\alpha}-E_{\alpha '}+p}
{\cal H}_{\alpha \alpha' l_n L L'L^*L'^*}
 - \int\frac{dp\,p}{E_{\beta}+E_{\alpha '}+p}
{\cal H}_{\beta \alpha'l_nLL'L^*L'^*}
 \right ] \biggr \} ,
\end{equation}
where we define function
\[
{\cal H}_{\alpha\alpha'l_nLL'L^*L'^*}= {\cal
H}_{\alpha\alpha'l_nLL'L^*L'^*}(p)
\]
\begin{equation}
= H^2_{\alpha \alpha 'l_n}\delta_{lLL^*}\delta_{l'^{\pm}L'L'^*} +
H^2_{\alpha ' \alpha l_n}\delta_{l^{\pm}LL^*}\delta_{l'L'L'^*} -
H_{\alpha \alpha 'l_n} H_{\alpha ' \alpha l_n} (\delta_{lL}
\delta_{l^{\pm}L^*} \delta_{l'^{\pm}L'} \delta_{l'L'^*}
+\delta_{lL^*} \delta_{l^{\pm}L} \delta_{l'L'} \delta_{l'^{\pm}
L'^*} )
\end{equation}
and radial integrals
\[
 H_{\alpha \alpha 'l_n}= H_{\alpha \alpha 'l_n}(p)=
 \int dr\left[r^2 f_{\alpha'}(r)g_{\alpha}(r)j_{l_n}(pr)\right],
\]
\[
R_{\alpha \alpha 'l_n}= R_{\alpha \alpha 'l_n}(p)=
 \int dr\left[r^2 g_{\alpha'}(r)g_{\alpha}(r)j_{l_n}(pr)\right],
\]
\begin{equation}
F_{\alpha \alpha 'l_n}= F_{\alpha \alpha 'l_n}(p)=
 \int dr\left[r^2 f_{\alpha'}(r)f_{\alpha}(r)j_{l_n}(pr)\right].
\end{equation}

The angular momentum coefficients ${\cal A}, \, {\cal B}, \, {\cal
D}$ and ${\cal E}$ can be found from  Appendix C of Ref.
\cite{tur09}.
\subsection{ Exchange diagrams contribution}
 \par
\begin{figure}[tbh]
\begin{center}
\includegraphics[width=15cm]{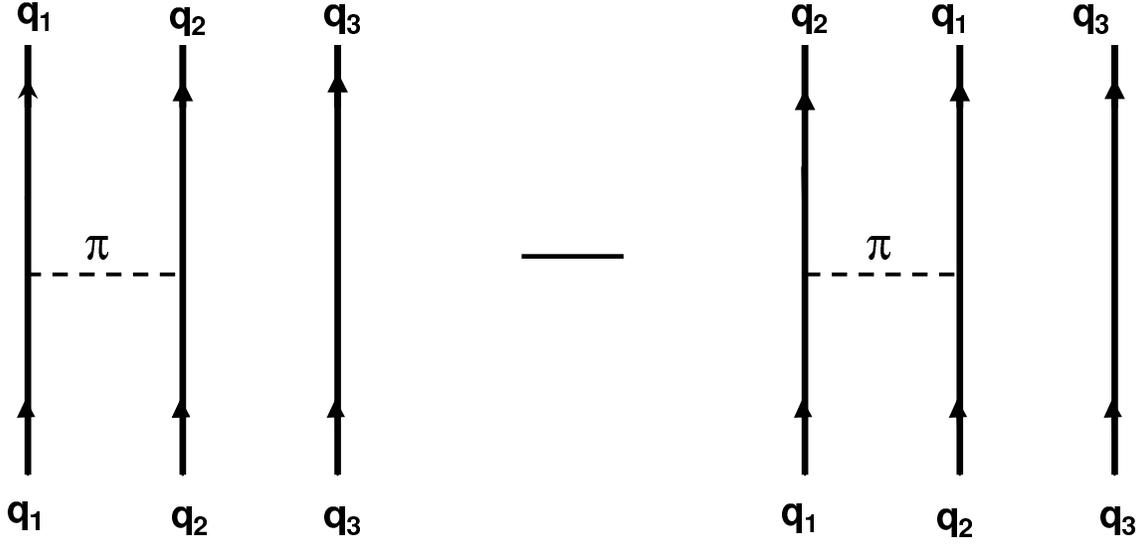}
\end{center}
\caption{Second order $\pi-$meson exchange diagrams
 \label{Fig2a}}
\end{figure}

 The pion exchange contribution to the baryon energy-shift (see Fig.\ref{Fig2a} ) is evaluated as:
 \begin{eqnarray}
\Delta
E_{ex.}^{(\pi)}=-\frac{1}{2f_{\pi}^2}\sum\limits_{a=1}^{3}\sum\limits_{\alpha
 \leq  \alpha_F}\sum\limits_{\alpha ' \leq \alpha_F} \int\frac{d^3\vec p}
{(2\pi)^3p_0^2} \biggl\{ V_{\alpha \alpha  }^{a+}(\vec p) V_{\alpha
'\alpha ' }^{a}(\vec p)- V_{\alpha \alpha '}^{a+}(\vec p) V_{\alpha
\alpha '}^{a}(\vec p) \biggr\}.
\end{eqnarray}
 By using the Wick's theorem we can write a more convenient  expression for the
 energy shift of the SU(2) baryons from the  second order pion exchange
 diagrams:
  \begin{eqnarray}
\Delta E_{ex.}^{(\pi)}=-\frac{1}{16\pi^3f_{\pi}^2}\int \frac{d p \,
p^2}{p_0^2} \sum\limits_{l_n}\Pi_{l_n}(p)
\end{eqnarray}
where
\begin{eqnarray}
\label{pionex}
 \Pi_{l_n}(p)=<\Phi_B|\sum\limits_{i\neq j}\vec
\tau(i)\vec\tau(j)T_{ln}(i) T_{l_n}(j)K_{l_n}(i)K_{l_n}^+(j)|\Phi_B>
\,
\end{eqnarray}
and the operators $\vec \tau, T_{l_n}$ and $ K_{l_n}$ are summed
over single quark levels $i\neq j$ of the SU(2) baryon. In the quark
model, the baryon wave function $ |\Phi_B> $ is presented as a bound
state of three valence quarks
 in the orbital configuration $(1S)^2(nlj)$,
and it can be written down commonly as
\begin{eqnarray}
\nonumber |\Phi_B>=|\alpha\beta\gamma>(J_0T_0)=
|\alpha\beta;\gamma>_{JM(J_0)}^{TM_T(T_0)} \\
\nonumber =\hat {S}\biggl [ |\psi_{\alpha}(r_1)\psi_{\beta}(r_2)
\psi_{\gamma}(r_3){\cal Y}_{J_0}^{JM}(\hat{x_1}\hat{x_2};\hat{x_3})>
|\chi_{T_0}^{TM_T}(12;3)>\biggr ]|\chi_c^(123)>,
\label{baryonstate}
\end{eqnarray}
where $J_0$ and $T_0$ are intermediate spin and isospin couplings of
the two S-wave valence quarks, respectively. They satisfy the
symmetry requirement $S_0=T_0$.  The states $\psi$ are the single
particle states, labeled by a set of quantum numbers $\alpha$,
$\beta$ and $\gamma$, excluding the color degree of freedom.

\par  The operator $T_{l_n}$ in equation (\ref{pionex}) is the
radial integration operator:
\begin{eqnarray}
<\alpha|T_{l_n}|\beta>=\int dr\biggl [r^2
S(r)j_{l_n}(pr)G_{\alpha\beta}(r)\biggr ].
\end{eqnarray}
with
\begin{eqnarray}
  G_{\alpha \alpha '}(r)=f_{\alpha}(r)g_{\alpha '}(r) + f_{\alpha '}(r)
g_{\alpha }(r)  .
\end{eqnarray}
 where $\alpha=(N,l,j,m_j,m_t,m_c)$ and $\alpha^{\prime}$ are
 two sets of the single quark quantum numbers.
 The matrix elements of the operator $K_{l_n}$ are given by
\begin{eqnarray}
\label{pion}
 \nonumber <\alpha|K_{l_n}|\beta>=-\biggl( 4\pi
[l^{\pm}(\alpha)][l_n][j(\alpha)]\biggr)^{1/2}C^{l(\beta)0}_{l^{\pm}
(\alpha)0l_n 0} \\
 W (j(\alpha)\frac{1}{2}l_n,l(\beta); l^{\pm(\alpha)},j(\beta))
C^{j(\beta)m(\beta)}_{j(\alpha)m_j(\alpha)l_n(m(\beta)-m(\alpha))},
\end{eqnarray}
and the Hermitian conjunction
\begin{equation}
\nonumber
 <\alpha|K_{l_n}^+|\beta>=<\beta|K_{l_n}|\alpha>,
\end{equation}
where $j(\alpha), l(\alpha), l^{\pm}(\alpha), m(\alpha)$ are the
quantum numbers of the single quark state $<\alpha|$.

\par
\begin{figure}[tbh]
\begin{center}
\includegraphics[width=15cm]{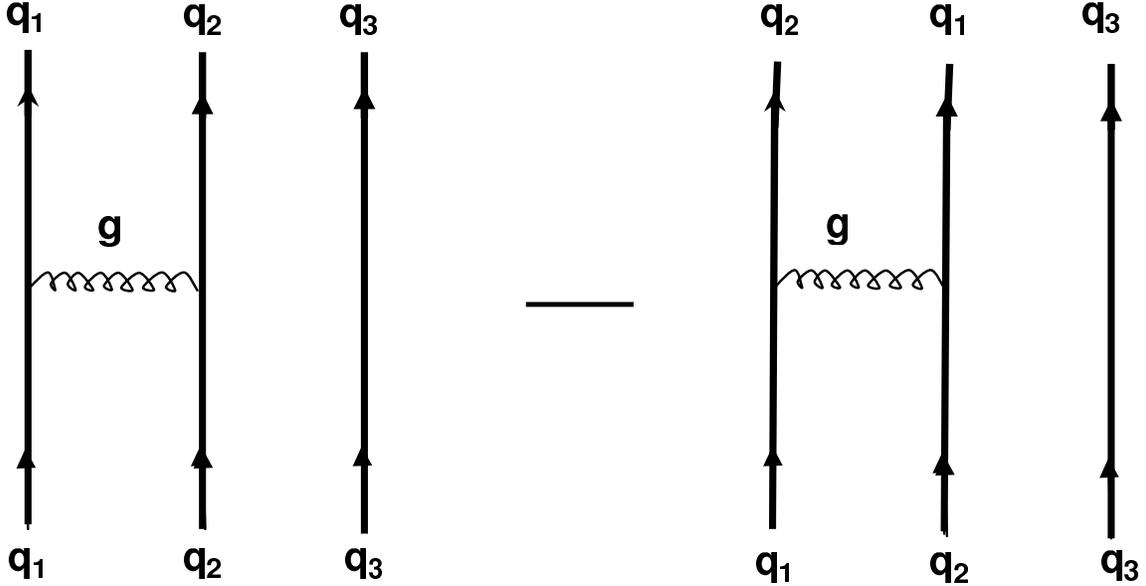}
\end{center}
\caption{Second order gluon-exchange diagrams
 \label{Fig2b}}
\end{figure}

 The contribution of the second-order gluon-exchange terms to the baryon spectrum
 (see Fig.\ref{Fig2b}) is given by
 \begin{eqnarray}
 \Delta E_{ex.}^{(g)}=-\frac{g^2}{2}\sum\limits_{a \mu
\nu}\sum\limits_{\alpha
  \leq \alpha_F}\sum\limits_{\alpha ' \leq \alpha_F} \int\frac{d^3\vec p}
{(2\pi)^3p^2} \biggl \{ V_{\alpha \alpha  }^{a\mu+}(\vec p)
V_{\alpha '\alpha ' }^{a\nu}(\vec p)- V_{\alpha \alpha
'}^{a\mu+}(\vec p) V_{\alpha  \alpha '}^{a\nu}(\vec p) \biggr \}
g^{\mu\nu}.
\end{eqnarray}
By using the Wick's theorem we can write  more convenient expression
for this equation
  \begin{eqnarray}
 \Delta E_{ex.}^{(g)}=-\frac{g^2}{\pi}\int
\limits_0^{\infty}dp
 \sum\limits_{l_n m_n} {\cal Q}_{l_n m_n}(p)
\end{eqnarray}
with the corresponding color-electric (Coulomb) and color-magnetic
parts:
\begin{eqnarray}
\label{gluon}
 \nonumber {\cal Q}_{l_n m_n}(p)
=<\Phi_B|\sum\limits_{i \neq j}\frac{\vec \lambda(i)}{2} \frac{\vec
\lambda(j)}{2}T^{(g)}_{ln}(i)T^{(g)}_{ln}(j)
\hat F_{l_n m_n}(i) \hat F_{l_n m_n}^+(j) |\Phi_B> \,  \\
-<\Phi_B|\sum\limits_{i \neq j}\frac{\vec \lambda(i)}{2} \frac{\vec
\lambda(j)}{2}T^{(g)}_{ln}(i)T^{(g)}_{ln}(j) \hat F_{l_n m_n}(i)
\hat F_{l_n m_n}^+(j) \hat {\vec \alpha}(i) \hat {\vec
\alpha}(j)|\Phi_B> .
\end{eqnarray}
The operator $T^{(g)}_{ln}$ is the radial integration operator with
the
 factor $j_{l_n}(pr)$. The operators $\hat F_{l_n m_n}(i)$
and $\hat F^+_{l_n m_n}(j)$ are the angular integration operator
with the factors
 $Y_{l_nm_n}(\hat x_i)$ and $Y^*_{l_nm_n}(\hat x_j)$ respectively. All these
 operators are summed over single quark levels $i \neq j$ of the SU(2) baryon.
\subsection{Selection rules for the quantum numbers of the excited
 $N^*$ and $\Delta^*$ states}
\par Now we begin to analyze the excited
$N^*$ and $\Delta^*$ spectrum based on the relativistic description
of  one-pion and one-gluon (color magnetic part) exchange
mechanisms. These exchange operators, as was found in
Ref.\cite{tur09}, couple the upper and lower components of the two
interacting valence quarks, respectively. Based on this fact we can
derive the selection rules for the quantum numbers of the baryon
states with the fixed orbital configuration.
\par Let us to fix the orbital configuration as $(1S_{1/2})^2(nlj)$,
with the intermediate spin coupling $\vec{S}_0=\vec{S}_1+\vec{S}_2 =
\vec{1/2}+\vec{1/2}$ of the two $1S$-valence quarks, where the last
valence quark $(nlj)$ can be in the ground or an excited state. The
upper and lower Dirac components of the last excited valence quark
have orbital momenta $l$ and $l' =l\pm 1$, respectively. Our choice
of the above orbital configuration is close to the limitation in the
diquark-quark models \cite{diquark}, where some of the degrees of
freedom are "frozen". The corresponding baryon states are different
from members of the SU(6)$\otimes$O(3) multiplets in the Constituent
Quark Models.

\par The first two selection rules come from the coupling of the three
valence quarks into the SU(2) baryon state with total momentum $J$
and isospin $T$:
\begin{eqnarray}
\label{quarkcoup}
 \nonumber
\vec{S}_0+\vec{j}=\vec{J} \\
 \nonumber
 \vec{T}_0+\vec{1/2} =\vec{T} \\
   T_0 =S_0
\end{eqnarray}
where the symmetry property of the two S-quarks coupling was used.
The third rule comes from the pion exchange mechanism between the
excited valence quark and the $1S$ quark. This mechanism couples the
upper (lower) component of the $1S$ valence quark with the lower
(upper) component of the excited $(nlj)$ valence quark. Since the
upper component of the S-quark has zero orbital momentum, then for
the orbital momentum of the exchanged pion we derive the equation
\begin{equation}
\label{piorbit}
 L_{\pi}=l'=l \pm 1
\end{equation}

\par   The final selection rule is based on the assumption that the coupling of
the last valence quark with quantum numbers $(nlj)$ to the $1S$
quark plus pion  is the main component of the strong coupling of the
excited baryon state to the $N(939)+\pi$:
\begin{equation}
\label{piN}
 \vec{L}_{\pi}+\vec{1/2} =\vec{J}
\end{equation}
With this assumption, Eq.(\ref{piorbit}) can be used for the
identification of the baryon resonance in the $\pi N$-scattering
process. Namely, when $l'=0$ we have S-wave nucleon and delta
resonances, when $l'=1$ we have P-wave resonances, etc.
\par An important consequence of the obtained selection rules is
that all the $N^*$ and $\Delta^*$ resonances appearing in the $\pi
N$ scattering process and coupling strongly to the $\pi N$ channel
are identified with the orbital configurations $(1S_{1/2})^2(nlj)$
with two valence quarks in the ground state and a single valence
quark in an excited state. A baryon resonance corresponding to the
orbital configuration with two valence quarks in excited states
$(1S_{1/2})(nlj)_1(nlj)_2$ couples strongly to the $\pi \pi
N$-channel, but not to the $\pi N$ channel.

\par Using the obtained selection rules it is very natural to
analyze schematically the excited nucleon and delta spectrum. For
the fixed orbital configurations $(1S_{1/2})^2(nlj)$ with the
intermediate spin coupling of the two $S-$ wave quarks $S_0=0$ (the
so-called instanton channel), Eq.(\ref{quarkcoup}) allows only a
single $N^*$ state with $J=j$ and no any $\Delta^*$ resonances.
\par Except the case, when the last valence quark is in the $P_{1/2}$ orbit,
the intermediate coupling $S_0=1$, due-to the selection rule
Eq.(\ref{piN}) yields two resonances in the both nucleon and delta
sectors with the total momentum $J=L_{\pi} \pm 1/2$. In this way one
of the $N^*$ resonances defined by the selection rules in
Eq.(\ref{quarkcoup}) with $J=j+1$ or $J=j-1$ is ruled out. When the
last valence quark is in the $P_{1/2}$ orbit, i.e. has the lower
S-component, the selection rules yield $L_{\pi}=0$ and $J=1/2$, and
consequently, only single S-wave resonances in the both nucleon and
$\Delta$ sectors are allowed.
\par Thus, for the fixed $(1S_{1/2})^2(nlj)$  orbital configuration
with $(nlj) \ne (nP_{1/2})$ there must be a band of three $N^*$ and
two $\Delta^*$ resonances. The lightest $N^*$ state corresponds to
the intermediate spin coupling $S_0=0$ due-to strong attraction in
this "instanton channel". The other two $N^*$, as well as the two
$\Delta^*$ resonances correspond to the spin coupling $S_0=1$ and
must be close each to other.
\par In the case when the last quark is
in the $P_{1/2}$ orbit, there is a band of two $N^*$ states (not
close each to other) and a single $\Delta^*$ resonance appearing in
the S-wave of the $\pi N$ scattering.

\section{Numerical results}
\subsection{Condition of the calculations}

 \par In order to account for the finite size effect of the pion, we introduce
a one-pion vertex regularization function in the momentum space,
parameterized in the dipole form as
$$ F_{\pi}(p^2)= \frac {\Lambda_{\pi}^2-m_{\pi}^2}{\Lambda_{\pi}^2+p^2}.$$
We fix $\Lambda_{\pi}=1$ GeV in our calculations. Contrary to the
bag-model calculations, the above regularization is used not for the
solution of the convergence problem of the quark self-energy. This
was explicitly shown in Ref.\cite{gutsche89} and \cite{tur10} for
the lowest valence quark states. As is known from
Ref.\cite{saito84}, the convergence of the quark self-energy is a
serious problem in the bag models.
\par As was noted above, the strength
$c=0.16$ GeV$^2$ and Coulomb  $\alpha=\pi/12\approx 0.26$ parameters
 of the Cornell potential are fixed from the flux-tube study
\cite{lus81} and lattice calculations \cite{kawa11,green03}.
However, it is useful to note that the above value of the strength
parameter was already probed long times ago in Ref.\cite{gutsche87}.
The only free parameter of the model, $m$ of the confining potential
was chosen as m=60 MeV  to reproduce the correct axial charge of the
proton $g_A=1.26$ (and the empirical pion-nucleon coupling constant
$G^2_{\pi NN}/4\pi=14$ via the Goldberger-Treiman relation). It
yields a reasonable value for the quark core RMS  radius of the
proton 0.52 fm (see \cite{gutsche87}). The strong coupling constant
$g_s^2=4\pi\alpha_s$ with the value $\alpha_s=0.65$.

\par In Ref. \cite{gutsche87} by examining the different  model parameters
the sensitivity of the Nucleon energy on the description of the
static properties of the proton has been examined. It was found that
a larger value of the strength parameter $c$ of the confining
potential yields a smaller value for the proton RMS radius.

\par Stating that the Coulomb like term of the Cornell potential
$ V(r)=-\alpha/r $ is actually due-to the color electric component
of one-gluon exchange mechanism, we need to avoid a double counting
of these components in the calculations of gluon loop corrections to
the baryon mass spectrum. This is why we have restricted our study
to the color-magnetic component of the one-gluon exchange forces
together with one-pion loop corrections.

\par In Ref.\cite{tur10} we have demonstrated explicitly a convergence of
the self-energy for the valence quarks in the lowest $1S$, $2S$,
$1P_{1/2}$, $1P_{3/2}$ orbits induced by the pion and color-magnetic
gluon fields. The total momentum of the intermediate quark and
anti-quark states increases from $j=1/2$ up to $j=25/2$, while their
radial quantum numbers grows up to $n=20$ in order to reach
convergent results.
\par We have obtained convergence of the self-energy also for the excited
valence quark states in the orbits $3S$, $2P_{1/2}$, $2P_{3/2}$,
$1D_{5/2}$, which are included into the structure $(1S)^2(nlj)$ of
the excited baryons in present study.
\par By summing the self energies of the three valence quarks in the
excited $(1S)^2(nlj)$ nucleon and Delta states, we can estimate the
contribution of the self-energy terms to the excitation spectrum of
the SU(2) flavor baryons.

\subsection{Ground state Nucleon spectrum}

\par In Table 1 we give the mass values for the g.s. N(939) with and
without CM correction in three different methods: the  R=0,
\cite{lu98},  P=0 \cite{teg82} and  LHO \cite{wil89}. All these
methods were firstly examined in Ref. \cite{shimizu}. As we can  see
from the Table, they agree within 50 MeV for the ground state
Nucleon.
\par The pion loop diagrams yield positive contribution to the baryon
mass-spectrum due-to self-energy term. For the ground state Nucleon
it is 200 MeV.
\par For the gluon field contributions we probe two different ways.
In the first case we include the contribution of all the
intermediate quark and antiquark states up to convergence with
$j=$25/2. The corresponding results are given in the 3-row of Table
1, they increases the Nucleon mass by 109 MeV. In the second case  a
restriction of  the intermediate states to the ground $1S$ quark
state is used when estimating the self-energy (I=0). The second
approximation is based on the short-range character of one-gluon
exchange forces. The corresponding energy shift for the ground state
Nucleon is now negative (-127 MeV). However, after including the
center of mass corrections, the Nucleon mass is still overestimated
by about 100 MeV. In principle, we can fit the strong coupling
constant $\alpha_s$ to reproduce the N(939) mass value, but first we
have to check the excitation spectrum of the SU(2) flavor baryons.
Thus, from the results in Table 1 we can conclude, that the second
way, when the short-range character of one-gluon exchange forces is
taken into account, is most favorable.
\par We note that the agreement within 50 MeV of the three
 R=0 \cite{lu98},  P=0 \cite{teg82} and  LHO \cite{wil89} methods for the
CM correction is reasonable.  Moreover, these three methods always
give corrections with systematic differences. Namely, the LHO method
always yields correction larger than the $P=0$ method, but smaller
than the $R=0$ method (see Ref. \cite{tur09}).
 Thus, we can fix one of these methods (R=0) and go to the excited sector.

\subsection{Spectrum of the SU(2) flavor baryons}

\par In Table 2 we compare our numerical estimations of the excited
$N^*$ and $\Delta^*$ spectrum within the developed schematic
periodic  table with the last experimental data from \cite{rev} and
\cite{anis}. The calculations were done up-to and including F-wave
baryon resonances in the frame of the developed chiral quark model.
In the Table we give the center of mass (CM) corrected quark core
results (zero order estimation) (second column) together with the
second order pion field contributions corresponding to the self
energy (3-th column) and exchange diagrams (4-th column). In order
to reproduce the ground state Nucleon and Delta quark core energy
value, the parameter of the effective Coulomb-like vector potential
for the diquark in Eq.(\ref{effecpot}) is slightly modified:
\begin{eqnarray}\label{effecpot1}
 \hat V (\vec r) =\frac{5 \alpha} {r},
\end{eqnarray}
while keeping other parameters of effective potentials in
Eq.(\ref{effecpot}) as before. In this way the scalar diquark energy
value decreases from 632 MeV to the reasonable value of 520 MeV as
estimated with the help of Eq.(\ref{twobody}). The estimated
$1S_{1/2}$ single quark energy value is 420 MeV  as found from the
solution of Eq.(\ref{single}) with the modified potentials from
Eq.(\ref{effecpot}), that yields for the total quark-core energy of
the ground state Nucleon an estimation 940 MeV, consistent with the
results of the $R=0$ method (see Table 1). At the end, by using
these effective potentials we have estimated the quark-core energy
values of the excited $N^*$ and $\Delta^*$ resonances on the basis
of the developed in the Section \ref{CMmodel} method.
\par The 5-th column of the Table 2 contains results for the quark core
plus pion loop corrections. Next 6-th and 7-th columns correspond to
the contributions of the self-energy and exchange terms of the
color-magnetic one-loop diagrams. The final theoretical estimations
are given in the 8-column with the strong coupling constant
$\alpha_s=0.65$. As was argued above, due to the short range
character of the gluon exchange forces between valence quarks, we
restrict our calculations of the color-magnetic self-energy terms to
the case, where the intermediate quark is the same initial and final
quark.
\par Based on obtained selection rules first we will show the assignment
of the excited baryon states presented in the data from
Ref.\cite{rev} with corresponding orbital configurations. Let us to
fix the orbital configuration $(1S_{1/2})^2(nS_{1/2})$. In the data
there are four $N^*$ with $J^{\pi}=1/2^+$ ( $P_{11}$ resonances) and
two $N^*$ with $J^{\pi}=3/2^+$ ( $P_{13}$ resonances). With the
above rules, we can find easily that $N^*(1440)$,  $N^*(1710)$, and
$N^*(1720)$ resonances belong to the orbital configuration
$(1S_{1/2})^2 (2S_{1/2})$ with the radially excited $2S$ valence
quark state, while the other three $N^*(1880)$, $N^*(1900)$ and
$N^*(2100)$ resonances correspond to the orbital configuration
 $(1S_{1/2})^2(3S_{1/2})$. In the $\Delta$ sector there are two
resonances with $J^{\pi}=3/2 ^+$  at  1600 MeV and 1920 MeV, and two
states with $J^{\pi}=1/2 ^+$ at 1750 MeV and 1910 MeV which belong
to the orbital configuration with the radially excited valence quark
in consistence with our results.
\par The orbital configuration $(1S_{1/2})^2(1D_{3/2})$ is not
presented in the data, since it would give two $N^*$ resonances with
$J^{\pi}=3/2^+$ and a single  $N^*$ resonance with $J^{\pi}=1/2^+$.
\par For the orbital configurations $(1S_{1/2})^2(nP_{1/2})$ there
are four nucleon and three delta resonances with $J^{\pi}=1/2 ^-$
and they are not close each to others. Each of the nucleon bands
$n=1$ and $n=2$  contains two resonances, while $\Delta^*$
resonances correspond to the three bands including $n=3$.
\par The orbital configuration $(1S_{1/2})^2(nP_{3/2})$ with $n=1$
 yields three $N^*$ resonances $3/2^-$(1520), $5/2^-$(1675) and $3/2^-$(1700),
 the first of which is less than other two states in accordance with our prediction.
The band with $n=2$ yields next group of the D-wave Nucleon
resonances  $3/2^-$(1860), $3/2^-$(2080) and $5/2^-$(2200).
 \par In the Delta sector there are four D-wave resonances, however
 only two of them $\Delta(5/2^-)$(1930) and  $\Delta(3/2^-)$(1940)
 are close each to other. Since other D-wave resonances
$\Delta(3/2^-)$(1700) and  $\Delta(5/2^-)$(2350) are far, then we
can predict possible new  $\Delta^*(5/2^-)$ (around 1700 MeV) and
$\Delta^*(3/2^-)$ (around 2350 MeV) resonances.
\par The F-wave $N^*$ resonances $N^*(5/2^+)(1680)$,
$N^*(5/2^+)(1870)$  and  $N^*(7/2^+)(1990)$ belong to the orbital
configuration $(1S_{1/2})^2(nD_{5/2})$ with $n=1$ together with
delta states $\Delta^*(5/2^+)$(1905) and $\Delta^*(7/2^+)$(1950),
while the $\Delta^*(5/2^+)$(2000) and $\Delta^*(7/2^+) $(2390)
belong to the $n=2$ band.
\par We can continue our analysis at higher energies and predict
in summary seven new $N^*$  resonances  with $J^{\pi}=7/2^-$ ~ (2000
MeV), $9/2^+$ ~(2100 - 2300 MeV),~
 $11/2^+$ (2100 - 2300 MeV),~ $11/2^-$ ~(2500-2700 MeV),~ $13/2^-$ (2500-2700 MeV),
 ~$13/2^+$ ~(2600 -2800 MeV), ~$15/2^+$ ~(2600 -2800 MeV) and four  $\Delta^*$
 resonances with $J^{\pi}=5/2^-$~ (around 1700 MeV),~ $3/2^-$ ~(2350 MeV),
 ~$11/2^-$ ~(2750 MeV),~ $13/2^+$ ~(2950 MeV). These resonances are expected to be observed
 in current experimental facilities.
\par It is clear now that the remaining "missing $N^*$ and $\Delta^*$ resonances"
predicted by the Constituent Quark Models must appear in the $\pi
\pi N$ strong coupling sector, if they exist. As we have argued
above, they will be assigned with the orbital configuration
$(1S_{1/2})(nlj)_1(nlj)_2$ with two excited valence quarks and a
single ground state valence quark.
\par
\begin{figure}[tbh]
\begin{center}
\includegraphics[width=20cm]{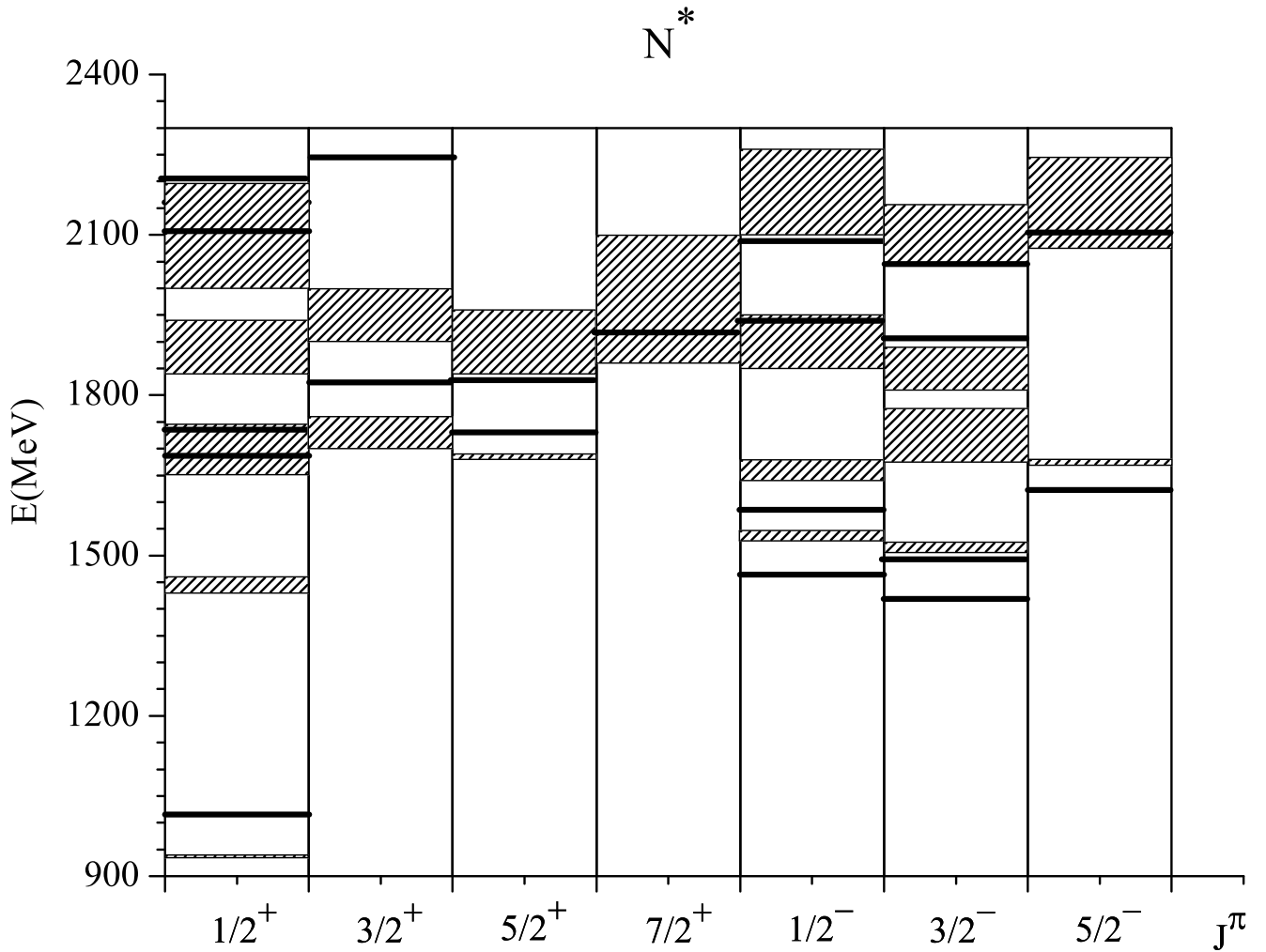}
\end{center}
\caption{Spectrum of the Nucleon states. Theoretical estimations
(solid lines) in comparison with
 experimental data (boxes)
 from Ref.\cite{rev}
 \label{Fig1}}
\end{figure}

\begin{figure}[tbh]
\begin{center}
\includegraphics[width=20cm]{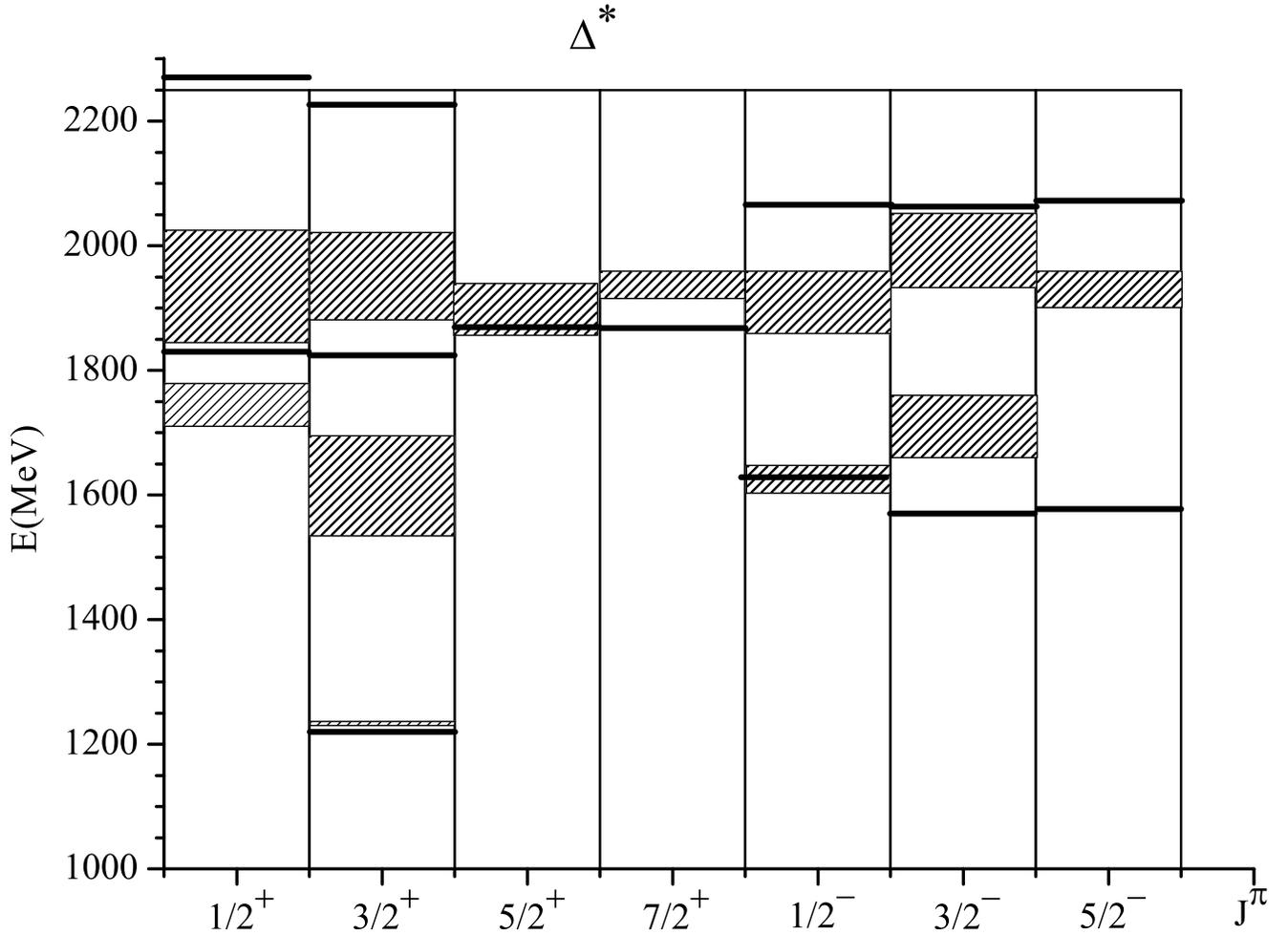}
\end{center}
\caption{Spectrum of the Delta states (notations are the same as in
 Fig.\ref{Fig1} )
 \label{Fig2}}
\end{figure}

  Now we can analyze the numerical values within our model
in comparison with the experimental data from Ref. \cite{rev}. In
Fig.\ref{Fig1} and Fig.\ref{Fig2} we give the theoretical
estimations and experimental data in a convenient diagrammatic way.
Table 2 contains information about orbital configurations for each
baryon resonance, as well as separate contributions from self-energy
and exchange diagrams due-to pion- and color-magnetic gluon fields.
As can be seen from the Table and figures, the mass spectrum of the
Nucleon and $\Delta$ is described reasonably well in the
relativistic chiral quark model with a single free parameter of the
confining potential.
\par For the test of the results we can check the consistence of our
results with the results of the Cloudy Bag Model \cite{thomas80}.
The pion exchange diagrams contribute about 144 MeV to the energy
difference between $N(939)$ and $\Delta(1232)$, while the gluon
exchange forces yield 64 MeV for the strong coupling constant value
$\alpha_s=0.65$. The value $\alpha_s=1.51$ increases the gluon field
contribution up to 149 MeV, which is consistent with the CBM
results. However, as one can see from the Table, this way strongly
moves down almost all the baryon states including  $N(939)$ and
$\Delta(1232)$.
\par The next important observation is that one needs an
additional exchange mechanism for the lowering the ground state
$N(939)$ and resonances $N(1440)$ (Roper), $N^*(1720)(3/2^+)$,
$N^*(1880)(1/2^+)$ and $N^*(1900)(3/2^+)$. On the other hand, two of
the radially excited nucleon resonances, $N^*(1710)(1/2^+)$ and
$N^*(2100)(1/2^+)$  are inside the corresponding error boxes.
 \par  The close situation is in the  $\Delta$ sector. The ground state
 $\Delta(1212)$  is well reproduced. However, the first radial excitation band is
 slightly overestimated ($\Delta(1600)3/2^+$ and  $\Delta(1750)1/2^+$), while the second
radial excitation band is overestimated strongly.

\par Contrary, the first band of orbitally excited $N^*$ resonances with a negative parity are
mostly underestimated. The second band is inside or close to the
experimental box. The situation in the $\Delta^* $ sector is close.
The orbitally excited $\Delta^*$ states corresponding to the lowest
radial quanta $n=1$  are slightly underestimated or inside the
experimental box, while negative parity $\Delta$ states
corresponding to the radial quantum number n=2 are mostly
overestimated.

\par The orbitally excited nucleon and delta resonances with the
positive  parity are reproduced quite well in the developed model.

\par It is relevant to compare the obtained estimations for the excited
N* an d $\Delta*$ spectrum with the results of the relativized
Constituent Quark Model \cite{cap00}. A comparison of  the results
presented in the Fig.\ref{Fig1} and Fig.\ref{Fig2} with the results
presented in Fig.9 and Fig.10 of above mentioned work indicates that
the two methods describe the excited baryon spectrum approximately
at the same level. However, the present model does not have any
fitting parameters, and, additionally, unlike CQM,  it does not
predict many missing nonobserved resonances.

\par The analysis shows that one needs an additional exchange
mechanism between valence quarks to reproduce the whole SU(2) baryon
spectrum. The new exchange forces  must depend on the spin and
flavor of valence quarks as well as on the quantum numbers of the
baryon state. Of course, large part of the interaction comes from
two-pion exchange mechanism.

\section{Conclusions}
 \par In summary, we have derived selection rules for the excited baryon state, assuming
 that it's orbital configuration is of the form $(1S)^2(nlj)$ with two valence
 quarks in the ground state and a single excited quark. These selection rules were derived
 on the basis of the one-pion exchange mechanism between valence
 quarks in the frame of the relativistic chiral quark model.
 An important consequence of the obtained selection rules is that all
the $N^*$ and $\Delta^*$ resonances appearing in the $\pi N$
scattering process and strongly coupled to the $\pi N$ channel are
identified with the orbital configurations $(1S_{1/2})^2(nlj)$.
 Baryon resonances corresponding to the orbital configuration with
two valence quarks in excited states couple strongly to the $\pi \pi
N$-channel, but not to the $\pi N$ channel.

\par Based on obtained selection rules, we have constructed a schematic periodic table
and calculate the energy spectrum of the excited $N^*$ and
$\Delta^*$ baryons within the field-theoretical framework including
one-pion and one-gluon loop corrections. The zero-order energy
values of the SU(2) flavor baryons are estimated including the
Center of mass corrections in a new method, based on the separation
of the three-quark core Hamiltonian into three parts, corresponding
to the Jacobi coordinates. The obtained numerical estimations for
the energy positions of baryon resonances (up to and including
F-wave) yield an overall good description of the experimental data.
However, Nucleon ground state and most of the radially excited
Nucleon resonances (including Roper) are overestimated. Contrary,
the first band of the orbitally excited $N^*$ resonances with a
negative parity are underestimated, while the second band is close
to the experimental boxes. The positive parity nucleon resonances
with J=5/2+ and 7/2+ are within or close to the experimental boxes.
In the $\Delta $ sector we have a similar situation, however, the
second excitation band ($n=2$) of the orbitally excited $\Delta$
states with a negative parity are mostly overestimated. At the same
time, the ground state $\Delta(1232)$ is well reproduced.
\par The important observation is that one needs an additional exchange
mechanism for the lowering both the ground state $N(939)$ and the
radially excited  $N^*$ and $\Delta^*$ resonances, including the
Roper resonance $N(1440)$. Of course, the two-pion exchange forces
are expected to contribute essentially to the excited baryon
spectrum.

\par A comparison of  the obtained results  with the results of the
relativized Constituent Quark Model indicates that they describe the
excited baryon spectrum approximately at the same level. This level
of description in our model was achieved without any fitting
parameters. Moreover, unlike CQM, our model does not yield many
nonobserved resonances at the lower excitation spectrum. The only
$\Delta(5/2^-)$ resonance is expected to be observed at energy scale
around 1600-1800 MeV.
\par At higher energies, where the experimental data are poor, we can
extend our model schematically and  predict the existence of seven
new $N^*$ and four $\Delta^*$ states with larger spin values. Of
course, the number of "missing resonances" in our model is strongly
suppressed due-to restriction of the configuration space to the
orbits $(1S_{1/2})^2(nlj)$. However, as we have shown above, at
lower energies this construction works reasonably well.

\section*{Acknowledgments}
\par One of the authors (E.M.T.) thanks Th. Gutsche for his valuable advices and discussions,
 A. Rakhimov for useful discussions, and K. Shimizu for the help in
the calculations of the center-of mass corrections for the ground
state Nucleon and Delta baryons. His work was supported in part by
the DAAD (Germany) Research Fellowship Programm. He acknowledges the
Institute fuer Kernphysik, Forschungszentrum Juelich, Germany for
the kind hospitality during his research stay.

\begin{table}
\caption { The mass value of the g.s. nucleon in MeV  with and
without center of mass (CM) correction}
 \begin{tabular}{|c|c|c|c|c|c|}     \hline
    & No CM   & R=0, \cite{lu98}& P=0, \cite{teg82}& LHO, \cite{wil89}   \\    \hline
  $E_Q$                     & 1715    & 940    &985     & 966      \\
  $E_Q+\Delta E(\pi)$       & 1915    & 1140   &1185    &1166      \\
  $E_Q+\Delta E(\pi+g)$     & 2024    & 1249   &1294    &1275      \\
  $E_Q+\Delta E(\pi+g)$, I=0& 1788    & 1013   &1058    &1039     \\ \hline

\end{tabular}
\end{table}

\begin{table}
\caption { Estimations for the energy values of the  $N^*$ and
$\Delta^*$ resonances in MeV }
 \begin{tabular}{|c|c|c|c|c|c|c|c|c|}     \hline
 SU(2) baryon state& $E_Q(CMcor.)$&$ \Delta E_{\pi}^{s.e.}$ &$\Delta E_{\pi}^{ex.}$ &
$E_Q+\Delta E_{\pi}$ & $\Delta E_g^{s.e.}$ &$ \Delta E_g^{ex.}$ &
E(theor) & E(exp.)\cite{rev}  \\  \hline

$N(939)(1/2^+)(P_{11})$ $(1S)^3$  & 940 & 380 & -180 & 1140 & -95 &
-32 & 1013 & 938 $\div$ 939 \\
$N(1440)(1/2^+)(P_{11})$ $(1S)^2(2S)$ & 1289 & 603 &-113 & 1750&
-70& -24 &
1685 & 1430 $\div$ 1470 \\
$N(1710)(1/2^+)(P_{11})$ $(1S)^2(2S)$ & 1289 & 603 &-66 & 1797 &-70 &-10 &1746 & 1650 $\div$ 1750  \\
$N(1720)(3/2^+)(P_{13})$ $(1S)^2(2S)$ & 1289 & 603 & 1 & 1864  &-70 &10 &1833 & 1700 $\div$ 1760  \\

$N(1880)(1/2^+)(P_{11})$ $(1S)^2(3S)$ & 1528 & 788 &-110 & 2166&-66 &-28&2112& 1840 $\div$ 1940  \\

$N(2100)(1/2^+)(P_{11})$ $(1S)^2(3S)$ & 1528 & 788 &-39 & 2237 &-66 &-1 &2210 & 2000 $\div$ 2200  \\

$N(1900)(3/2^+)(P_{13})$ $(1S)^2(3S)$ & 1528 & 788 &-3  & 2273 &-66 & 11 &2256& 1900 $\div$ 2000  \\

 $N(1535)(1/2^-)(S_{11})$ $(1S)^2 1P_{1/2}$ & 1186&501&-119&1541 &-79 &-13 &1476&1528$\div$1548
 \\
$N(1650)(1/2^-)(S_{11})$ $(1S)^2 1P_{1/2}$ & 1186&501&46 &1706
&-79&-49&1605& 1640$\div$1680  \\
 $N(1905)(1/2^-)(S_{11})$ $(1S)^2 2P_{1/2}$&1440&713&-111&2004&-69 &-27&1946& 1850$\div$1950
  \\
 $N(2090)(1/2^-)(S_{11})$ $(1S)^2 2P_{1/2}$ & 1440&713&24&2139 &-69&-13 &2095&  2100$\div$2260
 \\
 $N(1520)(3/2^-)(D_{13})$ $(1S)^2 1P_{3/2}$ & 1165 &515 & -126 & 1508 &-91&-27&1436&1518 $\div$ 1526  \\
$N(1700)(3/2^-)(D_{13})$ $(1S)^2 1P_{3/2}$ & 1165 &515 & -79 &
1555&-91 &-9
&1501& 1675 $\div$ 1775  \\
$N(1675)(5/2^-)(D_{15})$ $(1S)^2 1P_{3/2}$ & 1165 &515 & 11 & 1645 &-91 &29 &1629&1670 $\div$ 1680 \\
$N(1860)(3/2^-)(D_{13})$ $(1S)^2 2P_{3/2}$ & 1437 &713 & -111 &1983
&-73 &-29&1937& 1810 $\div$ 1890  \\

$N(2080)(3/2^-)(D_{13})$ $(1S)^2 2P_{3/2}$ & 1437 &713 & -31 &
2063&-73&
-1&2045&2045 $\div$ 2155  \\

$N(2200)(5/2^-)(D_{15})$
$(1S)^2 2P_{3/2}$ & 1437 &713 & 4 & 2098 & -73 &20&2101&2075 $\div$ 2245  \\
$N(1680)(5/2^+)(F_{15})$
$(1S)^2 1D_{5/2}$ & 1324 &638 & -114 & 1785 &-89&-30 &1729&1680 $\div$ 1690  \\
$N(1870)(5/2^+)(F_{15})$
$(1S)^2 1D_{5/2}$ & 1324 &638 & -37 & 1862& -89 & 2 &1838&1840 $\div$ 1960  \\
$N(1990)(7/2^+)(F_{17})$
$(1S)^2 1D_{5/2}$ & 1324 &638 & 12 & 1911 &-89 &27 &1912 &1860 $\div$ 2100  \\
$ \Delta(1232)(3/2^+)(P_{33})$  $(1S)^3$  & 940 & 380 & -36 & 1284
&-95 &32
&1221&1230 $\div$ 1234  \\
$ \Delta(1600)(3/2^+)(P_{33})$ $(1S)^2(2S)$ & 1289 & 603 & -23
&1841& -70&34 &1833 & 1535 $\div$ 1695  \\
 $ \Delta(1750)(1/2^+)(P_{31})$ $(1S)^2(2S)$ & 1289
& 603 & 1 &1865 & -70 &8 &1831& 1710 $\div$ 1780  \\
$\Delta(1910)(1/2^+)(P_{31})$ $(1S)^2(3S)$ & 1528 & 788 & -3 &2273 &
-66 &23&2270 & 1845 $\div$ 2025  \\
 $\Delta(1920)(3/2^+)(P_{33})$ $(1S)^2(3S)$ & 1528 & 788 & -18 &2258 & -66& 8 &2240 & 1880
$\div$ 2020  \\
$\Delta(1620)(1/2^-)(S_{31})$  $(1S)^2 1P_{1/2}$ & 1186&501& -24
&1636 &-79&45 &1629& 1603 $\div$ 1649  \\
 $\Delta(1900)(1/2^-)(S_{31})$  $(1S)^2 2P_{1/2}$ & 1440&713&  -24 &
 2091& -69& 12 &2072& 1860 $\div$ 1960  \\
$\Delta(1700)(3/2^-)(D_{33})$ $(1S)^2 1P_{3/2}$ & 1165 &515 &-18 &
1616 &-91& 8 &1579& 1670 $\div$ 1770   \\
  $\Delta(5/2^-)(D_{35})$ $(1S)^2 1P_{3/2}$ & 1165 &515 &-35 & 1599 & -91 &35 &1589 & ...   \\
$\Delta(1940)(3/2^-)(D_{33})$ $(1S)^2 2P_{3/2}$ & 1437 &713 &-9 &
2085 &-73&9 &2077& 1935 $\div$ 2055   \\
 $\Delta(1930)(5/2^-)(D_{35})$ $(1S)^2 2P_{3/2}$ & 1437 &713 &-22 & 2072 & -73& 28 &2083&1900$\div$ 1960   \\
$\Delta(1905)(5/2^+)(F_{35})$
$(1S)^2 1D_{5/2}$ & 1324 &638 & -12 & 1887 &-89 &7 &1868& 1860 $\div$ 1940  \\
$\Delta(1950)(7/2^+)(F_{37})$
$(1S)^2 1D_{5/2}$ & 1324 &638 & -27 & 1872 &-89 &29 &1875 & 1915 $\div$ 1960  \\

\end{tabular}
\end{table}

\end{document}